# The Logic of Experimental Tests, Particularly of Everettian Quantum Theory

David Deutsch

The Clarendon Laboratory, University of Oxford
david.deutsch@qubit.org

June 2016

Claims that the standard procedure for testing scientific theories is inapplicable to Everettian quantum theory, and hence that the theory is untestable, are due to misconceptions about probability and about the logic of experimental testing. Refuting those claims by correcting those misconceptions leads to an improved theory of scientific methodology (based on Popper's) and testing, which allows various simplifications, notably the elimination of everything probabilistic from the methodology ('Bayesian' credences) and from fundamental physics (stochastic processes).

**1. Introduction**

A version of quantum theory that is universal (applies to all physical systems), and deterministic (no random numbers), was first proposed informally by Schrödinger (Bitbol 1996) and then independently and in detail by Everett (DeWitt & Graham 1973). Subsequent work to elaborate and refine it has resulted in what is often called *Everettian quantum theory* (Wallace 2012). It exists in several variants, but I shall use the term to refer only to those that are universal and deterministic in the above senses. All of them agree with Everett's original theory that generically, when an experiment is observed to have a particular result, all the other possible results also occur and are observed simultaneously by other instances of the same observer who exist in physical reality – whose multiplicity the various Everettian theories refer to by terms such as 'multiverse', 'many universes', 'many histories' or even 'many minds[1]'.

Specifically, suppose that a quantum system **S** has been prepared in a way which, according to a proposed law of motion *L*, will have placed it in a state $|\psi\rangle$. An observable $\hat{X}$ of **S** is then perfectly measured by an observer **A** (where **A** includes an appropriate measuring apparatus and as much of the rest of the world as is affected). That means that the combined system $\mathbf{S} \oplus \mathbf{A}$ undergoes a process of the form

$$|\psi\rangle|0\rangle \longrightarrow \sum_i \langle x_i|\psi\rangle|x_i, a_i\rangle, \qquad (1)$$

where $|0\rangle$ denotes the initial state of **A**, the $|x_i\rangle$ are the eigenstates[2] of $\hat{X}$, and each $|a_i\rangle$ is a +1-eigenstate of the projector for **A**'s having observed the eigenvalue $x_i$ of $\hat{X}$. For generic $|\psi\rangle$, all the coefficients $\langle x_i|\psi\rangle$ in (1) are non-zero, and therefore all the possible measurement results $a_1, a_2, \ldots$ *happen*, according to *L* and Everettian quantum theory. Moreover, since interactions with the environment can never be perfectly eliminated, the measurement must be imperfect in practice, so we can drop the qualification 'generic': in all measurements, every possible result happens simultaneously.

Furthermore, the theory is deterministic: it says that the evolution of all quantities in nature is governed by differential equations (e.g. the Schrödinger or Heisenberg equations of motion) involving only those quantities and space and time, and thus it does not permit physically random processes. This is in contrast with certain other versions of quantum theory, and notably 'wave-function-collapse' variants in which measurements (and effectively only they) are governed by special, stochastic laws of motion whose effects are summarised in the *Born rule*. For perfect measurements, this is:

> If a system is initially in a state $|\psi\rangle$, and one of its observables $\hat{X}$ is measured, so that the state of the system+observer at the time $t$ immediately before completion of the measurement is $|\Psi(t)\rangle = \sum_i \langle x_i|\psi\rangle |x_i, a_i\rangle$, then immediately after time $t$ that state changes ('collapses') randomly to one of the $|x_i, a_i\rangle$ (meaning that the eigenvalue $x_i$ of $\hat{X}$ is observed) with probability $|\langle x_i|\psi\rangle|^2$. (2)

---

[1] 'Many minds' theories themselves exist in several versions. The original one (Albert and Loewer 1988) is probabilistic and therefore does not count as Everettian in the sense defined here. It is also dualistic (involves minds not being physical objects, and obeying different laws from theirs) and therefore arguably not universal. Lockwood's (1996) version avoids both these defects and is Everettian, so its testability is vindicated in this paper.

[2] Here and throughout, I am treating the measured observables as having discrete eigenvalues. The notation would be more cumbersome, but the arguments of this paper no different, for continuous eigenvalues.

In contrast, according to Everettian quantum theory, nothing singles out one of the $a_i$ from the others: they all happen; every possible result is experienced, and at times greater than $t$ the state of the combined system continues to evolve smoothly and unitarily from $|\Psi(t)\rangle$. What, then, can the predictive content of the theory, and consequently of laws of motion that conform to it such as *L* above, possibly be? That is the central problem addressed in this paper.

In particular, although the coefficients $\langle x_i|\psi\rangle$ are different for the different instances of the observer, nothing about their values (except for which of them are exactly zero, which in practice none of them ever are) describes any experience of the observer at the time *t*. Only the $a_i$ do. Hence it has been argued (e.g. by Kent 2010) that under Everettian quantum theory the coefficients $\langle x_i|\psi\rangle$ are purely 'decorative' – i.e. predictively irrelevant – making it what I shall call a mere *everything-possible-happens* theory. That is to say, the only thing it asserts about the reality described by (1) is that all the values $x_1, x_2, \ldots$, each with its corresponding result $a_1, a_2, \ldots$, including its effects on the observer, are simultaneously realised physically. Thus it would be a multi-universe theory that predicts nothing about the results of experiments except that (each instance of) the observer will experience some result among the $a_1, a_2, \ldots$ – i.e. among all results that are possible according to *L*. The same predictions about the results of experiments are made by the corresponding *something-possible-happens* theory – namely a single-universe theory saying that exactly one of the pairs $(x_i, a_i)$ describes what will happen in reality (and therefore that exactly one of the $a_i$ will be experienced), but saying nothing about which one.

Kent concludes on the basis of this supposed predictive equivalence with an associated everything-possible-happens theory, that Everettian quantum theory is untestable. And that therefore attempted single-universe versions of quantum theory (such as 'collapse' theories and 'pilot-wave' theories), whatever their weaknesses in other respects, are preferable, since they, at least in some "domain of validity", predict the *probabilities* of the possible results.

In fact, as I shall argue (Section 3), everything-possible-happens theories are not inherently untestable; and in any case, Everettian quantum theory is not one of them (Section 8). This will involve correcting some widespread misconceptions about probability and about the logic of experimental testing in general, which I shall do in Sections 2-4, where I shall present a version of scientific methodology that makes no use of probability, credences or theory confirmation. Then, in section 5, I shall discuss the status of stochastic theories, and in section 6, how the theory of experimental errors can be framed in that non-probabilistic methodology. And in Sections 7 and 8 I shall discuss testability in collapse-endowed theories and Everettian quantum theory respectively.

## 2. Explanations of explicanda

Prevailing discussions (e.g. Greaves & Myrvold 2010, Dawid & Thébault 2014) of the testability of various versions of quantum theory have approached the matter indirectly, in terms of *support* or *confirmation* – asking how our *credence* (degree of belief) for a theory should be changed by experiencing results of experiments. However, experimental confirmation is a philosophically contentious concept. Notably, it is rejected root and branch by Popper (1959). I shall present an account of the nature and methodology of scientific testing that closely follows Popper's. It differs from his, if at all[1], by regarding fundamental science as *exclusively* explanatory. That is to say, I take a *scientific theory* to be a conjectured explanation[2] (explanatory theory) of some aspects of the physical world – the *explicanda* of the theory – that is testable (I shall elaborate what that means below) by observation and experiment. A scientific *explanation* is a statement of what is there in reality, and how it behaves and how that accounts for the explicanda. Neither confirmation nor credence nor 'inductive reasoning' (from observations to theories or to justifications of theories as true or probable) appear in this account. So in this view the problem described in Section 1 is about *testing* theories.

This contradicts the 'Bayesian' philosophy that rational credences obey the probability calculus and that science is a process of finding theories with high rational credences, given the observations. It also contradicts, for instance, instrumentalism and positivism, which identify a scientific theory with its predictions of the results of experiments, not with its explanations. My argument here, that Everettian quantum theory is testable, depends on regarding it as an explanatory theory, and on adopting an improved notion of experimental testing that takes account of that.

Scientific methodology, in this conception, is not about anyone's beliefs or disbeliefs. Rather, it assumes that someone has conjectured explanatory theories (though it says nothing about how to do that), and it requires those who know (i.e. are aware of) those theories and want to improve them, to attempt to locate specific flaws and deficiencies and to attempt to correct those by conjecturing new theories or modifications to existing theories. Explicanda in the sciences usually involve *appearances* of some sort (e.g. the perceived blueness of the sky). Theoretical matters can also be explicanda (e.g. that classical gravity and electrostatics both have an inverse-square force law), but those will not concern us here. Explanations of appearances typically account for them in terms of an unperceived, underlying reality (e.g. differential scattering of photons of different energies) that brings about those appearances (though not only them).

---

[1] By acknowledging Popper's priority I am neither claiming that the account of scientific methodology that I shall present here is entailed by Popper's philosophy of science nor, on the other hand, that it departs significantly from it. Still less am I claiming that Popper himself would have agreed with its application to quantum theory – indeed, he rejected Everettian quantum theory (or at least, the versions known at the time). Here I only wish to show how my account solves the problem stated in Section 1 and illuminates several other matters.

[2] In this paper I am concerned with fundamental theories, but in fact any theory, to be testable, must at least explain why a particular experiment would constitute a test, because it must explain which variables would be confounding factors and which wouldn't. So strictly explanationless predictions cannot exist. (See Deutsch 2011 and the discussion of predictive oracles in Deutsch 1997.)

In this paper I shall be concerned with the part of scientific methodology that deals with experimental testing. But note that experimental testing is not the primary method of finding fault with theories. The overwhelming majority of theories, or modifications to theories, that are consistent with existing evidence, are never tested by experiment: they are rejected as bad explanations. Experimental tests themselves are primarily about explanation too: they are precisely attempts to locate flaws in a theory by creating new explicanda of which the theory may turn out to be a bad explanation.

Let me distinguish here between a *bad* explanation and a *false* one. Which of a theory's assertions about an explicandum are false and which are true (i.e. correspond with the physical facts) is an objective and unchanging property of the theory (to the extent that it is unambiguous). But how bad or good an explanation is depends on how it engages with its explicanda and with other knowledge that happens to exist at the time, such as other explanations and recorded results of past experiments. An explanation is better the more it is constrained by the explicanda and by other good explanations[1], but we shall not need precise criteria here; we shall

---

[1] For an informal discussion of good explanation, see Deutsch (2011).

only need the following: that an explanation is bad (or worse than a rival or variant explanation) to the extent that…

(i)     it seems not to account for its explicanda; or

(ii)    it seems to conflict with explanations that are otherwise good; or

(iii)   it could easily be adapted to account for anything (so it explains nothing).

It follows that sometimes a good explanation may be less true than a bad one (i.e. its true assertions about reality may be subset of the latter's, and its false ones a superset). Since two theories may have overlapping explicanda, or be flawed in different respects, the relations 'truer' and 'better' are both only partial orderings of explanations. Nevertheless, the methodology of science is to seek out, and *apparently* to correct, *apparent* flaws, conflicts or deficiencies in our explanations (thus obtaining better explanations), in the hope that this will correct real flaws and deficiencies (thus providing truer explanations).

When we, via arguments or experiments, find an apparent flaw, conflict or inadequacy in our theories, that constitutes a *scientific problem* and the theories are *problematic* (but not necessarily *refuted* yet – see below). So scientific methodology consists of locating and then solving problems; but it does not prescribe how to do either. Both involve creative conjectures – ideas not prescribed by scientific methodology. Most conjectures are themselves errors, and there need not be a right error to make next. Accordingly, all decisions to modify or reject theories are tentative: they may be reversed by further argument or experimental results. And no such event as 'accepting' a theory, distinct from conjecturing it in the first place, ever happens (cf. Miller 2006).

Scientific methodology, in turn, does not (nor could it validly) provide criteria for accepting a theory. Conjecture, and the correction of apparent errors and deficiencies, are the only processes at work. And just as the objective of science isn't to find evidence that justifies theories as true or probable, so the objective of the methodology of science isn't to find rules which, if followed, are guaranteed, or likely, to identify true theories as true. There can be no such rules. A methodology is itself merely a (philosophical) theory – a convention, as Popper (1959) put it, actual or proposed – that has been conjectured to solve philosophical problems, and is subject to criticism for how well or badly it seems to do that[1]. There cannot be an argument that certifies it as true or probable, any more than there can for scientific theories.

In this view a scientific theory is *refuted* if it is not a good explanation but has a rival that is a good explanation with the same (or more) explicanda. So another consequence is that in the absence of a good rival explanation, an explanatory theory cannot be refuted by experiment: at most it can be made problematic. If only one good explanation is known, and an experimental result makes it problematic, that can motivate a research programme to replace it (or to replace some other theory). But so can a theoretical problem, a philosophical problem, a hunch, a wish – anything.

An important consequence of this explanatory conception of science is that experimental results consistent with a theory *T* do not constitute support for *T*. That is because they are merely explicanda. A new explicandum may make a theory more problematic, but it can never solve existing problems involving a theory (except by making rival theories problematic – see Section 3). The asymmetry between refutation (tentative) and support (non-existent) in scientific methodology is better understood in this way, by regarding theories as explanations, than through Popper's (*op. cit.*) own argument from the logic of predictions, appealing to what has been called the 'arrow of modus ponens'. Scientific theories are only approximately modelled as *propositions*, but they are precisely *explanations*.

---

[1] For example, Popper's theory was proposed in order to avoid the problem of induction, and the problem of infinite regress in seeking authority ('justification') for theories, among other problems.

I now define an objective notion, not referring to probabilities or 'expectation values', of what it means for a proposed experiment to be *expected* to have a result *x* under an explanatory theory *T*. It means that if the experiment were performed and did not result in *x*, *T* would become (more) problematic. Expectation is thus defined in terms of problems, and problems in terms of explanation, of which we shall need only the properties (i)-(iii). Note that expectations in this sense apply only to (some) physical events, not to the truth or falsity of propositions in general – and particularly not to scientific theories: if we have any expectation about those, it should be that even our best and most fundamental theories are false. For instance, since quantum theory and general relativity are inconsistent with each other, we know that at least one of them is false, presumably both, and since they are required to be testable explanations, one or both must be inadequate for some phenomena. Yet since there is currently no single rival theory with a good explanation for all the explicanda of either of them, we rightly expect their predictions to be borne out in any currently proposed experiment.

A *test* of a theory is an experiment whose result could make the theory problematic. A *crucial test* – the centrepiece of scientific experimentation – can, on this view, take place only when there are at least two good explanations of the same explicandum (good, that is, apart from the fact of each other's existence). Ideally it is an experiment such that every possible result will make all but one of those theories problematic, in which case the others will have been (tentatively) refuted.

It will suffice to confine attention to problems arising from tests of fundamental theories in physics. And of those problems, only the simplest will concern us, namely when an existing explanation apparently does not account for experimental results. This can happen when there seems either to be an *unexplained regularity* in the results (criterion (i) above), or an *irregularity* (i.e. an explanation's prediction not being borne out – criterion (ii) above). So, if the result of an experiment is predicted to be invariably $a_1$, but in successive trials it is actually $a_5, a_{29}, a_1, a_3 \ldots$, with no apparent pattern, that is an apparent irregularity. If it is $a_5, a_5, a_5, a_5 \ldots$, that is apparently both an apparent unexplained regularity and an irregularity. Scientific methodology in this conception does not specify how many instances constitute a regularity, nor what constitutes a pattern, nor how large a discrepancy constitutes a prediction apparently not being borne out (as distinct from being a mere experimental error – see Section 6). Sometimes conflicting opinions about these matters can be resolved by repeating the experiment, or by testing other assumptions about the apparatus, etc.. But in any case, the existence of a problem with a theory has little import besides, as I said, informing research programmes – *unless* both the new and the old explicanda are well explained by a rival theory. In that case the problem becomes grounds for considering the problematic theory tentatively refuted. Therefore, to meet criterion (iii) above, it must not be protected from such a refutation by declaring it *ad hoc* to be unproblematic. Instead any claim that its apparent flaws are not real must be made via scientific theories and judged as explanations in the same way as other theories.

In contrast, the traditional (inductivist) account of what happens when experiments raise a problem is in summary: that from an apparent unexplained regularity, we are supposed to 'induce' that the regularity is universal[1] (or, according to 'Bayesian' inductivism, to increase our credence for theories predicting that); while from an apparent irregularity, we are supposed to drop the theory that had predicted regularity (or to reduce our credence for it). Such procedures would neither necessitate nor yield any explanation. But scientific theories do not take the form of predictions that past experiments, if repeated, would have the same outcomes as before: they must, among other things, *imply* such predictions, but they *consist* of explanations.

In any experiment designed to test a scientific theory *T*, the prediction of the result expected under *T* also depends on other theories: *background knowledge,* including explanations of what the preparation of the experiment achieves, how the apparatus works, and the sources of error. Nothing about the unmet expectation dictates whether $T$ or any of those background-knowledge assumptions was at fault. Therefore there is no such thing as an experimental result logically contradicting[1] $T$, nor logically entailing a different 'credence' for $T$. But as I have said, an apparent failure of $T$'s prediction is merely a problem, so seeking an alternative to $T$ is merely one possible approach to solving it. And although there are always countless logically consistent options for which theory to reject, the number of *good explanations* known for an explicandum is always small. Things are going very well when there are as many as two, with perhaps the opportunity for a crucial test; more typically it is one or zero[2]. For instance, when neutrinos recently appeared to violate a prediction of general relativity by exceeding the speed of light, no good explanation involving new laws of physics was, in the event, created, and the only good explanation turned out to be that a particular optical cable had been poorly attached (Adam et al. 2012).

---

[1] E.g. Aristotle's definition of induction as "argument from the particular to the universal".

Note that even if $T$ is the culprit, merely replacing it by $\sim T$ cannot solve the problem, because the negation of an explanation (e.g. 'gravity is not due to the curvature of spacetime') is not itself an explanation. Again, at most, finding a good explanation that contradicts $T$ can become the aim of a research programme.

I shall now show that it is possible for an explanatory theory $T$ to be testable even by an experiment for which $T$ makes only everything-possible-happens predictions, and whose results, therefore (if $T$ designates them as possible), cannot contradict those predictions.

---

[1] That is known as the *Duhem–Quine thesis* (Quine 1960). It is true, and must be distinguished from the Duhem–Quine *problem*, which is the misconception that scientific progress is therefore impossible or problematic.

[2] One of the misconceptions underlying the so-called 'problem of induction' is that since there is always an infinity of predictive formulae matching any particular data, science must be chronically overwhelmed with theories, with too few ways to choose between them. And hence that scientific methodology must consist of rationales for selecting a favoured theory from the overabundance: 'the simplest', perhaps, or the 'least biased'. But a predictive formula is not an explanation, and good *explanations* are hard to come by.

## 3. Refuting theories by their failure to explain

Suppose for simplicity that two mutually inconsistent theories, *D* and *E,* are good explanations of a certain class of explicanda, including all known results of relevant experiments, with the only problematic thing about either of them being the other's existence. Suppose also that in regard to a particular proposed experiment, *E* makes only the everything-possible-happens prediction (my discussion will also hold if it is a something-possible-happens prediction) for results $a_1, a_2, \ldots$, while *D* predicts a particular result $a_1$. If the experiment is performed and the result $a_2$ is observed, then *D* (or more precisely, the combination of *D* and the background knowledge) becomes problematic, while neither *E* nor its combination with the same background knowledge is problematic any longer (provided that the explanation via experimental error would be bad – Section 6 below).

Observing the result $a_1$, on the other hand, would be consistent with the predictions of both *D* and *E*. Even so, it would be a new explicandum which, by criterion (i) above, would raise a problem for the *explanation E*, since *why* the result $a_1$ was observed *but the others weren't* would be explained by *D* but unexplained by *E*. Note that if it were not for the existence of *D*, the result $a_1$ would not make *E* problematic at all. (Nor would any result, and so there would be no methodological reason for doing the experiment at all.)

If the experiment is then repeated and the result $a_1$ is obtained each time, that is an apparent regularity in nature. Again by criterion (i), *E* then becomes a bad explanation while *D* becomes the only known good explanation for all known results of experiments. That is to say, *E* is refuted (provided, again, that experimental error is a bad explanation). Although *E* has never made a false prediction, it cannot account for the new explicandum (i.e. the repeated results $a_1$) that its rival *D* explains.

Again, all refutations are tentative. Regardless of how often the above experiment is repeated with result $a_1$ every time, it remains possible that *E* is true – in which case the existence of a different explanation *D* with more accurate predictions may be a coincidence. But coincidence by itself could 'explain' anything, so, absent additional explanatory details, it must be a bad explanation by criterion (iii) above. A good explanation of all relevant observations might be some *E&G*, where *G* is a good explanation for why the result must be $a_1$ when the experiment is carried out under these circumstances, and why other results could be obtained under different circumstances.

Thus it is possible for an explanatory theory to be refuted by experimental results that are consistent with its predictions. In particular, the everything-possible-happens interpretation of quantum theory, to which it has been claimed that Everettian quantum theory is equivalent, could be refuted in this way (provided, as always, that a suitable rival theory existed), and hence it is testable after all. Therefore the argument that Everettian quantum theory itself is untestable fails at its first step. But I shall show in Section 8 that it is in fact much more testable than any mere everything-possible-happens theory.

It follows that under *E*, the string of repeated results $a_1$ is *expected not to happen*, in the sense defined in Section 2, even though *E* asserts that, like every other sequence, it *will* happen (among other things). This is no contradiction. Being *expected* is a methodological attribute of a possible result (depending, for instance, on whether a good explanation for it exists) while *happening* is a factual one. What is at issue in this paper is not whether the properties 'expected not to happen' and 'will happen' are consistent but whether they can both follow from the same deterministic explanatory theory, in this case *E*, under a reasonable scientific methodology. And I have just shown that they can.

Note that the condition that *E* be explanatory is essential to the argument of this section, which depends on the criteria (i) and (iii) for being a bad explanation. Under philosophies of science that identify theories with their predictions, theories like *E* would indeed be untestable and would inform no expectations. So much the worse for those philosophies.

**4. The renunciation of authority**

The reader may have noticed that the methodology I am advocating is radically different *in purpose*, not only in substance, from that which is taken for granted in most studies of the "empirical viability" of Everettian quantum theory, and of scientific theories in general. That traditional role of methodology has been to provide (1) some form of *authority* for theories – such as confirmation of their truth, justification, probability, credence, reason for believing, reason for relying upon, or 'secure foundations'; and (2) rules for using experiment and observation to give theories such authority. I am adopting Popper's view (e.g. Popper 1960) that no such authority exists, nor is needed for anything in the practice or philosophy of science, and that the quest for it historically has been a mistake.

Consequently, readers who conceive of science in terms of such a quest may regard the arguments of this paper as an extended acknowledgement that Everettian quantum theory is indeed fundamentally flawed in its connection to experiment, since in their view I am denying – for all theories, not just this one – the very existence of the connection they are seeking. Similarly, many philosophers regard Popper's own claim to have 'solved the problem of induction' as absurd, since his philosophy neither explains how inductive reasoning provides such authority nor (given his claim that no such reasoning exists) provides an alternative account of scientific reasoning that does provide it. This is not the place to defend Popper in this regard (but see Popper 1959). I merely ask readers taking such positions to conclude, from this paper, that the testability of Everettian quantum theory is not an *additional* absurdity separate from that of the non-existence of confirmation, inductive reasoning, etc.

To that end, note that when a methodology has authority as its purpose, it cannot consistently allow much ambiguity in its rules or in the concepts (such as 'confirming instance', or 'probability') to which they refer, because if two scientists, using different interpretations of the concepts or rules, draw different conclusions from the same experimental results, those conclusions cannot possibly both have authority in the above senses. But the methodology I am advocating is that of requiring theories to be good explanations and seeking ways of exposing flaws and deficiencies in them. So its rules do not purport to be sources of authority but merely summarise our "history of learning how not to fool ourselves" (Feynman 1974). It is to be expected that people using those rules may sometimes 'expect' different experimental results or have different opinions about whether something is 'problematic'. Indeed, explanation itself cannot be defined unambiguously, because, for instance, new modes of explanation can always be invented (e.g. Darwin's new mode of explanation did not involve predicting future species from past ones). Disagreeing about what is problematic or what counts as an explanation will in general cause scientists to embark on different research projects, of which one or both *may*, if they seek it (there are no guarantees), provide evidence by both their standards that one or both of their theories are problematic. There is no methodology that can validly guarantee (or promise with some probability etc.) that following it will lead to truer theories – as demonstrated by countless arguments of which Quine's (loc. cit.) is one. But if one adopts this methodology for *trying* to eliminate flaws and deficiencies, then despite the opportunities for good-faith disagreements that criteria such as (i)-(iii) still allow, one *may* succeed in doing so.

## 5. The status of stochastic theories

A stochastic theory (in regard to a particular class of experiments) is like a something-possible-happens theory except that it makes an additional assertion: that a 'random' one of the possible results $a_1, a_2, \ldots$ happens, with probabilities $p_1, p_2, \ldots$ specified by the theory. The 'random' values are either given as initial conditions at the beginning of time (as in pilot-wave versions of quantum theory) or produced by 'random processes' in which a physical system 'chooses randomly' which of several possible continuations of its trajectory it will follow (as in 'collapse' theories). In this section I argue that such a theory cannot be an explanatory description of nature, and under what circumstances it can nevertheless be useful as an approximation or mathematical idealisation.

We have become accustomed to the idea of physical quantities taking 'random' values with each possible value having a 'probability'. But the use of that idea in fundamental explanations in physics is inherently flawed, because statements assigning probabilities to events, or asserting that the events are random, form a deductively closed system from which no factual statement (statement about what happens physically) about those events follows (Papineau 2002, 2010). For instance, one cannot identify probabilities with the actual frequencies in repeated experiments, because they do not equal them in any finite number of repeats, and infinitely repeated experiments do not occur. And in any case, no statement about frequencies in an infinite set implies anything about a finite subset – unless it is a 'typical' subset, but 'typical' is just another probabilistic concept, not a factual one, so that would be circular. Hence, notwithstanding that they are called 'probabilities', the $p_i$ in a stochastic theory would be purely decorative (and hence the theory would remain a mere something-possible-happens theory) were it not for a special *methodological* rule that is usually assumed implicitly. There are many ways of making it explicit, but those that refer to individual measurement outcomes (rather than to infinite sequences of them, which do not occur in nature) and conform to the probability calculus, all agree on this:

> If a theory attaches numbers $p_i$ to possible results $a_i$ of an experiment,
> and calls those numbers 'probabilities', and if, in one or more instances
> of the experiment, the observed *frequencies* of the $a_i$ differ significantly,     (3)
> according to some statistical test, from the $p_i$, then a scientific problem
> should be deemed to exist.

(Or on equivalent procedures, under philosophies that do not refer to 'problems'.) Every finite sequence fails *some* test for randomness, and if a statistical test is designed to fail, its failure does not create a new explicandum, and hence does not make any theory problematic. Therefore one must choose the statistical test in the rule (3) independently of the experiment's results (e.g. in advance of knowing them).

The rule (3) does not specify how many instances, nor which statistical test, nor at what 'significance level' the test should be applied. In the event of a disagreement about those matters, the experiment can be repeated until the proposed statistical tests all agree. They will also all agree even for a single experiment if one of the $p_i$ is sufficiently close to 1. (Indeed, one can regard repeated instances of an experiment as a single experiment for which one of the $p_i$ is close to 1.)

This is the rationale under which a stochastic theory's assertions about probabilities are 'tested'[1]. I shall henceforward place the term in quotation marks when referring to such predictions because such 'tests' are not tests under the methodology of science I am advocating. They depend unavoidably on following rule (3), and the crucial thing about that rule is that it is *methodological,* and therefore normative*:* it is about how experimenters should behave and think in response to certain events; it is not a supposed law of nature, nor is it any factual claim about what happens in nature (the explicanda), nor is it derived from one. So it should not, by criterion (i), appear in a (good) scientific explanation. Nor, on the other hand, could it be appended to the explanatory scientific methodology I am advocating, for then it would be purely ad hoc: scientific methodology should be about whether reality seems to conform to our explanations; there is a problem when it does not, and only then. And *one cannot make an explanation problematic merely by declaring it so.* Nor, therefore, can one make a theory testable merely by promising to deem certain experimental results, consistent with the theory, problematic. If such results are unexplained by a theory (as in the example in Section 3), then the theory is already problematic and there is no need for a special rule such as (3). But if not, the theory is not problematic, and no ad hoc methodological rule can make it so.

Yet rule (3) is tacitly assumed by the Born rule (2), and consequently by 'collapse' theories, and by pilot-wave theories, which, like all stochastic theories, depend both for their physical meaning and for their 'testability' on (3). As I said, their 'probabilities' are merely decorative without it.

---

[1] A stochastic theory may also make non-probabilistic predictions and be tested through them. In particular, every stochastic theory makes something-possible-happens assertions, and these may allow it to be tested in the manner of Section 3. I shall give an example of this in Section 8.

So, how is it that stochastic theories can be useful in practice? I shall show in Section 7 how 'collapse' variants of quantum theory can, even though they are ruled out as descriptions of nature by the above argument. But most useful stochastic theories take the form of unphysically idealised models whose logic is as follows. The explicandum is some process whose real defining property is awkward or intractable to express precisely (such as the fairness of a pair of dice and of how they are thrown, or the non-designedness of mutations in genes). One replaces that property by the mathematical property of randomness. This method of approximation can be useful only if there is a good explanation for why one can expect the intended purpose of the model to be unaffected by that replacement. In the case of dice, the intended purpose of fairness includes things like the outcomes being unpredictable by the players. One would argue, among other things, that this purpose is achieved because the manner in which they are thrown does not give the thrower an opportunity to determine the outcome, because that would require motor control far beyond the precision of which human muscles and nerves are capable; and therefore that any pattern in the results *that was meaningful in the game* would be an unexplained regularity. Then one would argue that the optimal strategy for playing a game involving dice thrown in a realistic way is identical to *what it would be* in a game where the dice were replaced by a generator of random numbers – even though the latter is physically impossible. Then one can develop useful strategies without needing to calculate specifically what human hands and real dice would do, which would require intractable computations of microscopic processes even if one knew the precise initial conditions, of which one is necessarily ignorant.

But one thing that cannot be modelled by random numbers is ignorance itself (as in the 'ignorance interpretation of probability'). Fortunately, our ignorance of the microstate of the gas molecules in a flame has no bearing on why it warms us. The prediction (via the kinetic theory of gases, etc.) that it will radiate heat under particular circumstances rests not on ignorance but on a substantive, independently testable explanation of *why* it would make very little difference to regard the state of the flame as random with a particular probability distribution. That explanation is, again, that under the given circumstances any state of the flame that would *not* do so would have properties that would constitute an unexplained regularity and hence a new explicandum and a problem. So it is because we *know* something about the flame – not because of any of the things we do not know – that we expect it to warm us. Expectations, as always in science, are derived from explanatory theories.

The prevalence of the misconceptions about probability that I have described in this section has accustomed us to reinterpreting all human thought in probabilistic terms, as advocated by 'Bayesian' philosophy, but this is a mistake. When a jury is disagrees on what is 'beyond reasonable doubt' or 'true on the balance of probabilities', it is not because jurors have disagreed on the numerical value of some quantities obeying the probability calculus. What they have actually done (or should have done, according to the methodology I advocate here), is try to explain the evidence. Thus 'guilty beyond reasonable doubt' should mean that they consider all explanations that they can think of, that clear the defendant, *bad* by criteria such as (i)-(iii). And if they interpret those criteria differently, it should not be because they are with hindsight exploiting the leeway in them in order to deliver a particular verdict – because that would itself violate criterion (iii). Although the consequences of error may (or may not) be very different in the laboratory and the courtroom, what we should do – and all we can do, as I explained in Section 4 – is adopt methodologies, in science as in law and everything else, whose purpose is to facilitate error correction, not to create (a semblance of) authority for our surviving guesses.

## 6. Experimental error

Since error can never be perfectly eliminated from experiments, their results are meaningless unless accompanied by an estimate of the errors in them. Error estimates are often treated as probabilistic – either literally, in the sense of the results being treated as stochastic variables, or as credences. But since the approach I am advocating eschews both literal probabilities and credences, I must now give a non-probabilistic account of the nature of errors and error estimates.

Any proposed scientific experiment must come with an explanatory theory of how the apparatus, when used in practice, would work. This is used to estimate what I shall call the *least expected error* and the *greatest expected error*, which I shall define in a way that does not refer to probability.

For example, suppose that we use everyday equipment to measure out pieces of dough in a kitchen, intending them to have equal weights, and that later we find that they do indeed all have equal weights according to a state-of-the-art laboratory balance. The theories of that and of the kitchen apparatus will provide an explanation for why the latter is much less accurate (say, because of friction and play in its moving parts), so those results are unexpected under the definition of 'expect' in Section 2, and constitute an unexplained regularity (as with the dice in Section 4). They are therefore a new explicandum, and problematic by criterion (i). 'Coincidence' would be one explanation *consistent* with these events, but under the circumstances that would be a bad explanation by criterion (iii). The least error which, if subsequently discovered, would not make the theory of the apparatus problematic, I call the *least expected error*.

Similarly if, in the laboratory, one of the pieces was found to weigh more than the others combined, and if, according to our explanation of how the measuring-out process worked, such a piece would have been noticed during it and subdivided, that would be an unexplained *ir*regularity and therefore a problem by criteria (i) and (ii). The greatest such error which, if subsequently discovered, would not make the theory of the apparatus problematic, I call the *greatest expected error*.

Understood in this way, the greatest and least expected errors are properties of the state of explanatory theories about the apparatus and its use. Improving those theories could increase or reduce the expected errors even if the apparatus and its use are unchanged.

Experimental errors are often categorised as 'random' or 'systematic', but those terms can be misleading: Random errors are those caused by processes such as Brownian motion, which are well approximated by a *known* stochastic law. So they can be understood in the manner of Section 4. And in theory testing, where experiments are necessarily repeatable, such processes do not necessarily cause errors, because their effect could be made negligible just by repeating the experiment sufficiently often. So what remains to be explained here are systematic errors – that is, errors that are not random but *unknown*, and therefore conform to no known system. (This includes errors that could be approximated as random with suitable stochastic laws, perhaps depending on time and other conditions, if those were known.) Like the 'unknown unknowns' of military planning (Rumsfeld 2003), unknown variables in physics are counter-intuitive. Theories of them are something-possible-happens theories, which I have discussed, but what does it mean to *estimate* errors that are unknown, yet not random?

For example, consider measuring a real-valued physical quantity $\chi$ that all relevant explanations agree is a constant of nature – say, the speed of light in units of a standard rod and clock. For any given design of measuring instrument (say, that of Fizeau), even with repeated runs of the experiment whose results are optimally combined to give an overall result $x$, the greatest expected value of $|x-\chi|$ cannot be reduced beyond a certain limit, say $\varepsilon$, because of systematic errors. (I shall for simplicity consider only cases where the expectations of positive and negative errors are sufficiently similar for $|x-\chi|$ to be an appropriate quantity to estimate.)

As always with something-possible-happens theories, there will be a set of *possible* values, determined by various explanations. I have already mentioned the theories that $\chi$ is constant under suitable circumstances, and is a real number (not a vector, say). There will also be upper and lower bounds $a$ and $b$ on $x$, determined by explanations of the apparatus and of $\chi$ itself. For example, there was a limit on the speed at which Fizeau's cogged wheel could be smoothly rotated. And before the speed of light was ever measured, there was already a good explanation that it must be greater than that of sound because lightning arrives before thunder. Furthermore, even though $\chi$ has a continuous range of possible values, $x$ must be instantiated in a discrete variable, because even in cases where the result appears in a nominally continuous variable, like a pointer angle, it must always end up in a discrete form such as numerals recorded on paper, or in brains, before it can participate in the processes of scientific methodology. So $x$ has only finitely many possible values between $a$ and $b$. For simplicity, suppose that they have a constant spacing $\delta$.

So it would not make sense to estimate $|x-\chi|$ at less than $\delta/2$, nor of course at more than $b-a$. But what does it mean to estimate it at some $\varepsilon$ with $\delta \ll \varepsilon \ll b-a$? The estimate $\varepsilon$ cannot be regarded as probabilistic (e.g. as '$|x-\chi|$ is probably no more than $\varepsilon$'), since that would allow further measurements with the same instrument to improve the accuracy of the result, and we are dealing with the case where they cannot. Nor, for similar reasons, can it be regarded as a *bound* on $|x-\chi|$, since then, if $x_1$ and $x_2$ were two different results, we should have $|x_1-\chi|<\varepsilon$ and $|x_2-\chi|<\varepsilon$ and hence $\left|\frac{1}{2}(x_1+x_2)-\chi\right| < \varepsilon - \frac{1}{2}|x_1-x_2|$, which would contradict the assumption that $\varepsilon$ is the best estimate of the error. (Unless $x_1 = x_2$, but if all results were equal despite $\varepsilon \gg \delta$, this would be an unexplained regularity and a new problem, as in the previous example.) So the estimate can only mean, again as in that example, that if the true error $|x-\chi|$ were later revealed to exceed $\varepsilon$ (by a measurement that was more accurate according to an unrivalled good explanation), that would make our explanation of the *measurement process* problematic. And that explanation, being about unknown errors, cannot be of any help in increasing our knowledge of $\chi$: knowledge cannot be obtained from ignorance.

Some error processes have inherently bounded effects, but most explanations that inform error estimates are, like the ones in the dough example, not about the error processes themselves, but about processes in the experiment that are expected to prevent or correct or detect errors, but only if they are above a certain size. Hence, if we estimate $|x-\chi|$ at $\varepsilon$, and then perform the whole measurement several times with results $x_1, x_2, \ldots$, and $\max(|x_i - x_j|)$ turns out to be less than $\varepsilon$, that tells us nothing more about $\chi$ than we knew when we had performed it once. In particular, it need not (absent other knowledge) tell us that $\chi$ is 'likely' to be, or can be expected to be, between the largest and the smallest of the $x_i$. Indeed, as I said, if $\max(|x_i - x_j|)$ is *too* small (smaller than the least expected error), then the whole experiment is (absent further explanation) problematic, just as it is when some $|x_i - x_j|$ is greater than $\varepsilon$.

So both the inevitable presence in the multiverse of all possible errors after any experiment, and the validity of approximating some but not all of them as probabilistic, are consistent with the conception of science that I am advocating.

## 7. 'Collapse' variants of quantum theory

'Collapse' variants of quantum theory invoke a special law of motion, namely the Born rule (2), for times *t* when a measurement is completed. This law is unique in being the only stochastic law of motion ever proposed in fundamental physics[1]. Though vague, it does have a "domain of validity" that includes all experiments that are currently feasible. Within that domain, it allows those theories to be 'tested' as stochastic theories, according to the rationale described in Section 4, and all such 'tests' to date have been passed. Specifically, the logic of such a 'test' is as follows:

According to a law of motion $L$ under a 'collapse' theory with the Born rule, the probabilities of the possible results $a_1, a_2, \ldots$ of the given measurement are $p_1, p_2, \ldots$ respectively. A different law of motion (or a different version of quantum theory, or some other stochastic theory), predicts different probabilities $q_1, q_2, \ldots$. The two sets of probabilities follow from the only known good explanations of the phenomena in question. A statistical test is chosen, together with some significance level $\alpha$ and a number $N$, where $N$ is sufficient for the following. The experiment is performed $N$ times. Let $n_1, n_2, \ldots$ be the number of times that the result was $a_1, a_2, \ldots$ respectively. The statistical test distinguishes between three possibilities: (i) the probability is less than $\alpha$ that the observed frequencies $n_1/N, n_2/N, \ldots$ would differ by at least this much from the $p_1, p_2, \ldots$ and at least this little from the $q_1, q_2, \ldots$ if the probabilities really were $p_1, p_2, \ldots$; or (ii) the same with the *p*'s and *q*'s interchanged; or (iii) both. $N$ and $\alpha$ are chosen so as to make it impossible for neither (i) nor (ii) to hold[2]. If (iii) holds, the experiment is inconclusive. But if (i) holds, Rule (3) requires the experimenter behave as if the first theory had been (tentatively) refuted; or if (ii) holds, the same for the second theory. As I argued in Section 4, Rule (3) cannot make an explanation problematic; so there has to be an explanatory reason, not just a rule, for regarding the respective theories as problematic if that happens – see the following section.

---

[1] Pilot-wave theories and 'dynamical collapse' theories implement the Born rule indirectly, but are also stochastic theories in the sense I have defined. Though they do not give measurements a special status, the conclusions of this section apply to them too, because they depend on Rule (3) for their 'testing'.

[2] Confirmation-based methodologies would interpret (i) as 'second theory confirmed, first refuted' (to significance level α), and (ii) as vice-versa, and (iii) as 'neither theory confirmed, both refuted', and $N$ and α are chosen so as to make it impossible for both to be 'confirmed'.

For convenience in what follows, I shall re-state the statistical-test part of the above 'testing' procedure in terms of gambling: The experimenter considers a class of thought experiments in which two gamblers, one of whom knows only the first theory and the other only the second, bet with each other, at mutually agreed odds, on what each result $a_1, a_2, \ldots$, or combination of those results, will be. They behave rationally in the sense of classical (*probabilistic*) game theory, i.e. they try to maximise the expectation values of their winnings, as determined by the Born rule (2) and their respective theories. Since the $p_1, p_2, \ldots$ are not all equal to the corresponding $q_1, q_2, \ldots$, the experimenter can set odds for a pattern of bets on various propositions (such as 'the next outcome will be between $a_2$ and $a_7$'), to which each gambler would agree because each would calculate that the probabilistic expectation value of his winnings is positive for each bet. Each such pattern corresponds to a statistical test of whether the results are 'significantly' incompatible with the first theory, the second, or both. Finally, given the actual results of the experiments, the experimenter calculates the amount that the winner would have won (which is the amount that the loser would have lost). If it exceeds a certain value, rule (3) then requires the loser's theory to be deemed 'refuted' at the appropriate significance level. If neither wins more than that value, then the experiment is inconclusive.

Note, in passing, that these 'tests' of what their advocates call 'probabilistic predictions' are not the only ones by which theories framed under 'collapse' theories can be tested. They, like most versions of quantum theory, also make ordinary (i.e. non-probabilistic) predictions of, for instance, emission spectra, and the location of dark bands in interference patterns. It is true that in experimental practice there are small deviations from those predictions, for example due to differences between the real system and apparatus and idealised models of them assumed by the prediction, but in the absence of a rival explanatory theory predicting those deviations, they are validly treated as experimental errors, as in Section 6.

Note also that the Born rule's identification of the quantities[1] $|\langle x_i|\psi(t)\rangle|^2$ as probabilities cannot hold at general times *t* for any observable with eigenstates $|x_i\rangle$ for which interference is detectable, directly or indirectly. That is because during interference phenomena, i.e. almost all the time in almost all quantum systems, those quantities violate the axioms of the probability calculus (Deutsch et al. 2000).

## 8. Everettian quantum theory

Everettian quantum theory's combination of determinism with unpredictable outcomes of experiments has motivated two main criticisms of it in relation to probability. The one that I presented in Section 1, which denies the theory's testability (or misleadingly, its confirmability), is called the 'epistemic problem'. The other – the 'practical problem' – denies that *decisions* whose options have multiple simultaneous outcomes can be made rationally if all the possible outcomes of any option are going to happen anyway. The latter criticism has been rebutted by the so-called *decision-theoretic argument* (Deutsch 1999, Wallace 2003). Using a non-probabilistic version of game-theoretic rationality[2], it proves[3] that (in the terminology of the present paper) rational gamblers who knew Everettian quantum theory (and considered it a good explanation) but knew no 'collapse' variants of it (or considered them bad explanations), and have therefore made no probabilistic assumptions, when playing games in which randomisers were replaced by quantum measurements, would place their bets as if those were randomisers, i.e. using the probabilistic rule (2) according to the methodological rule (3).

---

[1] These are discrete quantities. For continuous eigenvalues *x*, the Born rule says that $|\langle x|\psi(t)\rangle|^2$, usually represented as $|\psi(x,t)|^2$, is a probability distribution function over *x*, at certain times *t*.

[2] For a comprehensive defence of that version of game-theoretic rationality, and rebuttals of counter-arguments, see Wallace (2012) Part II.

[3] A proof is not necessarily an explanation, so 'proved' in this context is a lesser claim than 'fully explained'. Quantum theory is currently in a similar explanatory state to that of general theory of relativity in its early years: it explained the phenomena of gravity decisively better than any rival theory, and from its principles many things could be proved about the orbits of planets, the behaviour of clocks, etc., including testable predictions; nevertheless the entity that it directly referred to, namely spacetime, was only imperfectly explained, so that, for instance, the event horizon in the Schwarzschild solution was mistaken for a physical singularity. Analogously, the entity that quantum theory directly refers to, namely the multiverse, is only imperfectly explained at present, in terms of approximative entities such as universes (Deutsch 2010).

The decision-theoretic argument, since it depends on game-theoretic axioms, which are normative, is itself a methodological theory, not a scientific one. And therefore, according to it, all valid uses of probability in decision-making are methodological too. They apply when, and only when, some emergent physical phenomena are well approximated as 'measurements', 'decisions' etc. so that the axioms of non-probabilistic game theory are applicable. Applying them is a substantive step that does not (and could not) follow from scientific theories.

Solving the 'practical problem' does not fully solve the 'epistemic problem', for one cannot directly translate the gamblers' situation into the scientist's. For instance, it is not clear how the scientist's equivalent of 'winnings' on discovering a true theory should be modelled, since there is no subjective difference between discovering a true theory and mistakenly thinking that one has. On the view, which I have argued here is a misconception, that scientific methodology is about generating confirmation of, or credence for, theories being true or probable, the 'epistemic' problem takes the form: 'the decision-theoretic argument proves that *if* one believes Everettian quantum theory, it is rational (in that non-probabilistic game-theoretic sense) to make choices as though the outcomes were probabilistic; but it cannot be rational in *any* sense to believe Everettian quantum theory in the first place because, by hypothesis, we would then believe that whenever some experimental evidence has led us to believe Everettian quantum theory, the contrary evidence was in reality present too'. There are arguments (e.g. Greaves 2007) that the 'epistemic' conclusion nevertheless follows from versions of the decision-theoretic argument, but unfortunately they share the same misconception about confirmation and credence, so for present purposes it would be invalid to appeal to them. Hence I must connect the decision-theoretic argument to theory testing via a route that does not assign confirmation, credence, or probability to theories.

The connection is, fortunately, straightforward: it is via *expectations*, which, in the non-probabilistic sense defined in Section 2, are not credences and do not obey the probability calculus. Nor, as I showed there, is it per se inconsistent to expect an experiment to have the result $x$ even though one knows that not-$x$ will also happen – provided that there is an explanation for such expectations, which, indeed, there is, as follows.

Suppose that gamblers who knew Everettian quantum theory and did not use any probabilistic rule such as (2) or (3), and were rational in the non-probabilistic sense required by the decision-theoretic argument, were to play the quantum-measurement-driven game of Section 7. That argument says that they would place exactly the same bets as described there. If one of them were to lose by a large enough margin, his expectation (in the non-probabilistic sense defined in Section 2) will have been violated, while the other's will not. Hence the loser's theory will be problematic and the winner's not. The experimenter who seeks good explanations can infer that if the gamblers were then informed of each other's theory, they would both consider that the loser's theory has been refuted, and hence the experimenter – who is now aware of the same evidence and theories as they are – must agree with them.

Following this testing procedure will (tentatively) refute different theories in different universes. As I said in Section 4, this is no defect in a methodology that does not purport to be guaranteed, nor probabilistically likely, to select true theories. However, note, as a reassuring consistency check (not a derivation – that would be circular!), that the decision-theoretic argument also implies that on the assumption that one of the theories is true, it is rational (by the criteria used in the argument) to bet that the other one will be refuted.

Consequently, the conventional modes of testing 'collapse' variants of quantum theory, or theories formulated under them, are valid for Everettian quantum theory: any experiment that 'tests' a probabilistic prediction of a 'collapse' variant is automatically also a *valid* test of the corresponding multi-universe prediction of Everettian quantum theory, because it does not depend on the Born rule nor any other assumption referring to probability.

Indeed, as I have argued in Sections 4 and 5, on this view of the logic of testability it is stochastic theories, including 'collapse' variants of quantum theory, that suffer from the very flaw that Everettian quantum theory is accused of having under conventional theories of probability, namely: conventional 'tests' of (or under) those variants all depend on arbitrary instructions such as the rule (3) about what experimenters should think. If it were deemed valid to add such instructions to a scientific theory's assertions about reality, the same could be done to Everettian quantum theory and the entire controversy about its testability would collapse by fiat. But it does not need special methodological rules. It is testable through its physical assertions alone.

So all scientific theories can (as they must) all be tested under the same methodological rules. There is no need, and no room, for special pleading on behalf of 'collapse' theories. Nor is there any room for stochastic theories, except those that can be explained as approximations in the sense of Section 4. Indeed, Albrecht & Phillips (2014) suggest that all stochastic phenomena currently known to physics are quantum phenomena in disguise.

Traditional 'collapse' theories are also inherently far worse explanations than Everettian quantum theory, by criterion (i), since they neither explain what happens physically between measurements, nor what happens during a 'collapse'. They also suffer from a class of inconsistencies known as the 'measurement problem' (thus failing criterion (ii)), which do not exist in Everettian quantum theory.

Note, however, that Everettian quantum theory is in principle testable against 'dynamical collapse' stochastic theories such as the Ghirardi–Rimini–Weber theory with fixed parameters, or any 'collapse' variants that specify explicitly enough the conditions under which 'collapse' is supposed to happen. That is because in principle one can then construct interference experiments that would produce a deterministic result $a_1$ under Everettian quantum theory but a range of possibilities $a_1, a_2, \ldots$ under those theories (Deutsch 1984). Those are something-possible-happens assertions, so even though the theories also assign 'probabilities' to those values, the something-possible-happens assertions alone, if borne out (with any result other than $a_1$), would be sufficient to refute Everettian quantum theory. If, in addition, repeated results had statistics close to those of the probabilistic assertions, there would be a new problem of explaining that: a literally stochastic theory is not explanatory, as explained in Section 5. If, on the other hand, repeated results $a_1$ were obtained, that would refute the non-Everettian theories, as explained in Section 3 – but they can also be ruled out without experimentation, as bad explanations.

## 9. Generalisation to constructor theory

Marletto (2015) has shown that if the decision-theoretic argument is valid, it also applies to a wide class of theories that conform to *constructor theory* (Deutsch & Marletto 2015). The arguments of this paper would apply to any theory in that class: they are all testable. (Everettian quantum theory is of course in the class; versions of quantum theory that invoke random physical quantities are not.)

## 10. Conclusions

By adopting a conception – based on Popper's – of scientific theories as conjectural and explanatory and rooted in problems (rather than being positivistic, instrumentalist and rooted in evidence), and a scientific methodology not involving induction, confirmation, probability or degrees of credence, and bearing in mind the decision-theoretic argument for betting-type decisions, we can eliminate the perceived problems about testing Everettian quantum theory and arrive at several simplifications of methodological issues in general.

'Bayesian' credences are eliminated from the methodology of science, but rational expectations are given an objective meaning independent of subjective beliefs.

The claim that standard methods of testing are invalid for Everettian quantum theory depends on adopting a positivist or instrumentalist view of what the theory is about. That claim evaporates, given that scientific theories are about explaining the physical world.

Even everything-possible-happens theories can be testable. But Everettian quantum theory is not one of them. Because of its explanatory structure (exploited by, for instance, the decision-theoretic argument) it is testable in all the standard ways. It is the predictions of its 'collapse' variants (and any theory predicting literally stochastic processes in nature) that are not genuinely testable: their 'tests' depend on scientists conforming to a rule of behaviour, and not solely on explanations conforming to reality.

**Acknowledgements**

I am grateful to David Miller, Liberty Fitz-Claridge, Daniela Frauchiger, Borzumehr Toloui and to the anonymous referees for helpful comments on earlier drafts of this paper, and especially to Chiara Marletto for illuminating conversations and incisive criticism at every stage of this work.

This work was supported in part by a grant from the Templeton World Charity Foundation. The opinions expressed in this paper are those of the author and do not necessarily reflect the views of Foundation.